\documentclass{article}

\usepackage{arxiv}

\usepackage[utf8]{inputenc} 
\usepackage[T1]{fontenc}    
\usepackage{hyperref}       
\hypersetup{colorlinks=true,allcolors=[rgb]{0,0,1}}
\usepackage{url}            
\usepackage{booktabs}       
\usepackage{amsfonts}       
\usepackage{nicefrac}       
\usepackage{microtype}      
\usepackage{lipsum}         
\usepackage{graphicx}
\usepackage{authblk}
\usepackage[table]{xcolor}

\usepackage{amsmath}

\usepackage{amssymb}
\usepackage{mathtools}
\usepackage{amsthm}
\usepackage{xspace}
\usepackage{multirow}
\usepackage{pgf}
\usepackage{tcolorbox}
\usepackage{booktabs}
\usepackage{boldline}
\usepackage{array}

\usepackage[round]{natbib}
\let\cite\citep

\usepackage[capitalize,noabbrev]{cleveref}

\usepackage{listings}
\usepackage[scaled=0.95]{inconsolata}

\usepackage{enumitem}

\usepackage[scaled=0.86]{helvet}

\title{PolicyLR: A Logic Representation for Privacy Policies}

\author{%
\textbf{Ashish Hooda}$^{1}$\quad \textbf{Rishabh Khandelwal}$^1$ \quad \textbf{Prasad Chalasani}$^2$\\\vspace{-8pt} \quad \textbf{Kassem Fawaz}$^1$ \quad
\textbf{Somesh Jha}$^{1}$\\\vspace{5pt}

$^1$University of Wisconsin-Madison \quad $^2$Langroid
}

\begin{document}
\maketitle

\newcommand{\name}{\lstinline{PolicyLR}\xspace}

\newcolumntype{?}{!{\vrule width 1pt}}

\definecolor{BaseRed}{rgb}{1,0,0}
\definecolor{BaseGreen}{rgb}{0,1,0}

\newcommand{\scalefactor}{50} 

\newcommand{\intensitycolor}[3]{
    \ifdim #1 pt > 0 pt
        \expandafter\cellcolor\expandafter{BaseRed!#2!white}#2 \%
    \else
        \expandafter\cellcolor\expandafter{BaseRed!#2!white}#2 \%
    \fi
}

\newcommand{\ratiocolor}[2]{
    \ifdim #1 pt > 0 pt
        \expandafter\cellcolor\expandafter{BaseRed!#2!white}#1 
    \else
        \expandafter\cellcolor\expandafter{BaseGreen!#2!white}#1 
    \fi
}
\begin{abstract}
Privacy policies are crucial in the online ecosystem, defining how services handle user data and adhere to regulations such as GDPR and CCPA. However, their complexity and frequent updates often make them difficult for stakeholders to understand and analyze. Current automated analysis methods, which utilize natural language processing, have limitations. They typically focus on individual tasks and fail to capture the full context of the policies. We propose \name, a new paradigm that offers a comprehensive machine-readable representation of privacy policies, serving as an all-in-one solution for multiple downstream tasks. \name converts privacy policies into a machine-readable format using valuations of atomic formulae, allowing for formal definitions of tasks like compliance and consistency. We have developed a compiler that transforms unstructured policy text into this format using off-the-shelf Large Language Models (LLMs). This compiler breaks down the transformation task into a two-stage translation and entailment procedure. This procedure considers the full context of the privacy policy to infer a complex formula, where each formula consists of simpler atomic formulae. The advantage of this model is that \name is interpretable by design and grounded in segments of the privacy policy. We evaluated the compiler using ToS;DR, a community-annotated privacy policy entailment dataset. Utilizing open-source LLMs, our compiler achieves precision and recall values of $0.91$ and $0.88$, respectively. Finally, we demonstrate the utility of \name in three privacy tasks: Policy Compliance, Inconsistency Detection, and Privacy Comparison Shopping.
\end{abstract}
\section{Introduction}

Privacy Policies play an integral part in the digital landscape by specifying how online services interact with users and their data. They provide a way to inform users how and why services collect, share, process, and retain their data. Service providers utilize privacy policies to show how their practices comply with applicable privacy norms and regulations, like the GDPR~\cite{gdpr} and CCPA~\cite{ccpa}. On the other hand, users can consult privacy policies to learn more about the practices of a service provider and make their decisions accordingly. For example, they can compare privacy policies of similar services to choose one that matches their privacy preferences.

However, privacy policies are long and opaque documents that users often struggle to read and comprehend~\cite{mcdonald2008cost,reidenberg2015disagreeable,pollach2007s, linden2018privacy}. Service providers regularly update their policies to comply with new regulations or reflect changes in their corresponding services, where one in five policies is updated every month~\cite{wagner2022privacy}. Coupled with the immense number of online services with which users interact, it is challenging for stakeholders, such as users and regulators, to keep track of up-to-date privacy policies.

Researchers leveraged the advances in natural language processing to automate the analysis and understanding of privacy policy documents through fine-grained labeling~\cite{harkous2018polisis}, summarization~\cite{zaeem2018privacycheck}, and question-answering-based approaches~\cite{ravichander2019question}. The automation of privacy policy analysis catalyzed progress in three main privacy policy tasks -- (1) structured representations for privacy policies, including alternative formats~\cite{cranor2002web}, nutrition labels~\cite{kelley2009nutrition,khandelwal2023overview}, privacy icons~\cite{regulation2016regulation}, and short notices~\cite{zimmeck2014privee}; (2) large-scale measurements to analyze the policy landscape at scale~\cite{amos2021privacy,wagner2022privacy}; and (3) mechanisms to find inconsistencies within a policy document~\cite{andow2019policylint} or verify compliance with privacy laws, like the GDPR~\cite{10172832,linden2018privacy, manandhar2024towards}.

These methods, however, suffer from two fundamental limitations. First, existing approaches are narrow in scope and only target individual tasks, either comprehension, consistency, or compliance. As such, they require training several task-specific ML models, making them hard to modify or extend. Second, they operate on a sentence level and fail to capture context across paragraphs. Due to the focus on sentence-level analysis, comprehension of policy text has been limited to one-dimensional attributes like data retention period or purpose. These approaches struggle to capture multidimensional concepts, such as \texttt{the retention period of location data in the context of European countries} which are likely to be described across multiple sentences or paragraphs of the policy document.

In this work, we propose \name, a logical representation of privacy policies, that promises to address limitations in existing approaches. \name represents a new paradigm of a comprehensive machine-readable representation that can serve as a one-for-all solution for multiple downstream tasks. We devise a logic system formulation to represent a privacy policy as the valuations of a set of atomic formulae. These formulae act as independent building blocks that can be combined together to convey multidimensional concepts like \texttt{All identifiable data used for advertisements should have a limited retention period}. The logic system allows for a formal definition of a variety of tasks such as compliance, consistency, and comparisons. 

To construct \name's logical representation, we first need to define a set of atomic formulae. We build upon existing work in policy annotation that has created hierarchical taxonomies of privacy concepts~\cite{wilson-etal-2016-creation,arora2022tale}. Specifically, we utilize the OPP-115 taxonomy developed by Wilson et al.~\cite{wilson-etal-2016-creation} to generate our list of atomic formulae. This taxonomy is widely recognized in the literature and offers a comprehensive set of privacy practices. Notably, the \name framework is designed to be flexible and independent of any specific taxonomy, allowing users to choose or customize taxonomies based on their requirements.

We build \name to be versatile in the sense that it can automatically generate its set of atomic formulae by ingesting existing machine-readable taxonomies for privacy policies. To make our logical representation compatible with existing privacy policies, we provide a compiler for \name that can transform any unstructured policy text into the valuations of a set of atomic formulae. Our compiler deconstructs the transformation task as a two-stage translation and entailment procedure. This formulation allows us to incorporate global context from different sections of the policy text. We use off-the-shelf instruction-tuned Large Language Models (LLMs) without any need for fine-tuning. To further improve its usability, we design our compiler to work with open-source LLMs.  

\name's formulaic representation comprises fundamental formulae, which are then used to construct more complex formulae. This atomics-based formulation helps in explaining the behavior of complex formulae. The entailment module provides valuations of these atomics along with evidence that cites relevant segments in the privacy policies that were used in the evaluation. This evidence helps ground LLMs' responses to relevant segments of the privacy policy. Both these features help make \name more interpretable and mitigate hallucinations.

\noindent \textbf{Contributions.} In this paper, we make the following contributions. 

\begin{enumerate}
    \item We propose \name, a logic-based representation for privacy policies. Our representation allows for formal definitions of various privacy tasks like compliance and consistency. We also provide an automated way to initialize \name's atomic formulae by leveraging existing privacy policy taxonomies. 
    \item We build a compiler that leverages open-source LLMs to transform an unstructured policy text into the valuations of a set of atomic formulae. \item We evaluate our compiler using ToS;DR, a community-annotated privacy policy entailment dataset. Our compiler achieves precision and recall values of $0.91$ and $0.88$ using open-source LLMs. 
    \item We demonstrate the utility of \name on three privacy tasks -- Policy Compliance, Inconsistency detection, and privacy comparison shopping.
\end{enumerate}
\section{Related Work}
We contextualize our proposed framework, \name, within related work around using LLMs for document reasoning, privacy policy analysis, document consistency analysis, and compliance analysis for policies.

\subsection{Reasoning about Documents Via Language Inference Task}
Large Language Models (LLMs) have emerged as powerful tools for natural language processing, demonstrating impressive abilities like text generation and comprehension~\cite{gao2023retrieval}. One crucial capability for reasoning about documents is \textit{Natural Language Inference} (NLI)~\cite{maccartney2009natural}. This task involves determining whether a natural language hypothesis can reasonably be inferred from a given premise. In the NLI task, LLMs are required to determine if the hypothesis is true (i.e., entailment), false (i.e., contradiction), or undetermined (neutral), given a premise. For example, consider a premise saying: ``\textit{John's brother is 8 years old.}''. A hypothesis that says ``\textit{John has no siblings.}'' will get a \textit{contradiction} label. \name uses LLM's NLI capabilities to map unstructured policy documents into valuations of a set of atomic formulae.

\subsection{Retrieval Augmented Generation for Document Reasoning}
While LLMs possess vast knowledge, they may not always have access to the specific information required for complex reasoning tasks involving specialized documents. Retrieval Augmented Generation (RAG)~\cite{lewis2020retrieval} addresses this limitation by combining the power of LLMs with external knowledge bases or document retrieval systems. In the context of document reasoning, RAG first retrieves relevant documents or passages from a corpus based on the given query or task. The retrieved information is then used to augment the LLM's input, providing contextual information and supporting evidence for reasoning. \name uses RAG to ground LLM reasoning on relevant parts of the privacy policy document.

\subsection{Privacy Policy Analysis}
Privacy policy analysis is crucial to understanding how organizations handle personal data. Automated analysis techniques are becoming increasingly important due to the vast number of privacy policies that exist. Early research focused on rule-based systems and supervised learning approaches for privacy policy analysis. Researchers have employed sentence classification approaches for tasks such as identifying missing information or categorizing policy elements~\cite{harkous2018polisis, bhatia2018semantic}. Cui et al. address semantic incompleteness through PoliGraph, a knowledge graph approach that analyzes entire policies using semantic role labeling~\cite{poligraph}.

Topic modeling techniques like those used by Sarne et al. can identify high-level themes within large collections of privacy policies~\cite{sarne2019unsupervised}. Shvartzshnaider et al.~\cite{shvartzshnaider2018analyzing} introduce a framework that analyzes policies from the perspective of information flow and user readability, considering contextual integrity.

Ontology-based techniques offer additional capabilities. PrivOnto by Oltramari et al. leverages an ontology to represent and analyze privacy policies, enabling semantic querying~\cite{oltramari2018privonto}. Nejad et al.'s Knight system focuses on mapping privacy policies to specific articles in regulations like GDPR~\cite{nejad2018knight}.

More recently, Large Language Models (LLMs) like ChatGPT and Llama 2 have shown significant promise for various NLP tasks~\cite{touvron2023llama}, including sentiment analysis and text summarization~\cite{el2021automatic}. Their ability to process complex language and identify patterns within large amounts of text makes them well-suited for extracting privacy practices from privacy policies. Rodriguez et al.~\cite{rodriguez2024large} propose using LLMs to extract privacy practices from the privacy policies. However, they use LLMs to perform zero-shot classification by treating all the classes as independent. 

These studies highlight the potential of NLP and semantic techniques for privacy policy analysis. However, most of these techniques rely on sentence-level analysis that does not account for the whole context in the document and can miss relationships between different sections of a policy. Furthermore, these techniques treat all classes as independent, e.g., \texttt{purpose} and \texttt{retention-period} classifiers are trained separately. This misses out on complex cases where these classes are not independent, and there is a need for a joint classification. \name accounts for these relations and represents privacy policy in a more granular and comprehensive manner.

\subsection{Document Consistency}
Document consistency is essential for ensuring the accuracy and clarity of information present within documents. Research in this area focuses on identifying inconsistencies within a single document (internal consistency) and across related documents (cross-document consistency).

\noindent\textbf{Intra-document Consistency.} Ali et al.~\cite{Ali2023Automatic} propose a system for automated consistency checks in financial documents by projecting the entities in embedding space and ensuring that semantic and syntactic variations are treated similarly. Andow et al. ~\cite{andowpolicylint} propose a system that analyzes privacy policies for internal inconsistencies by leveraging an ontology to capture positive and negative statements regarding data collection and sharing. These approaches segment the documents and use transformer-based models to classify segments that can miss out on the full context. \name, on the other hand, relies on state-of-the-art language models to understand the context and accurately determine privacy practices.

\noindent\textbf{Inter-document Consistency.} Researchers have also explored inter-document consistency to compare practices disclosed by developers in privacy labels with privacy policies ~\cite{khandelwal2023overview, jain2023atlas}. Jain et al.~\cite{jain2023atlas} introduce ATLAS, a system that casts consistency as a document classification task and automatically detects discrepancies between privacy policies and privacy labels for mobile apps. They extract privacy labels from privacy policies and compare them with actual labels released by the developers. Khandelwal et al.~\cite{khandelwal2023overview} create a new taxonomy for privacy labels and leverage it to build classifiers to predict privacy labels using privacy policies. Similarly, Zimmeck et al.~\cite{zimmeck2019maps} identify inconsistencies between the app's behavior and its stated privacy practices by combining machine learning-based privacy policy analysis with static code analysis. 

While previous research has made significant developments in consistency analysis, there are limitations. Sentence level analysis in PolicyLint~\cite{andowpolicylint} might miss additional context, resulting in erroneous results. Segment-level classification frameworks, as presented in Khandelwal et al.~\cite{khandelwal2023overview} and  Zimmeck et al.~\cite{zimmeck2016automated, zimmeck2019maps}, also suffer from this limited context problem. ATLAS~\cite{jain2023atlas}, on the other hand, poses the problem as document classification but trains 32 different privacy label-specific classifiers. \name, on the other hand, does not rely on segmentation and analyzes all relevant context. Furthermore, we implement \name using an off-the-shelf language model that acts as a universal compiler to generate the truth table, which in turn allows us to perform several downstream tasks.

\subsection{Policy Compliance}
Automated methods for assessing policy compliance with regulations are a growing area of interest. Manandhar et al.~\cite{manandhar2024towards} introduce the ARC framework for transforming complex privacy regulations into a structured format, facilitating automated analysis and compliance assessment. Prior works also focus on analyzing privacy policies for compliance with regulations like GDPR~\cite{linden2018privacy, liu2021have, liao2024understanding}. PolicyChecker by Liao et al. ~\cite{liao2024understanding} utilizes a rule and semantic role labeling approach to assessing compliance of mobile app privacy policies with GDPR. PTPDroid by Tan et al.~\cite{tan2023ptpdroid} identifies potential violations related to third-party data collection practices disclosed in Android app privacy policies. Shafei et al.~\cite{shafei2024measuring} investigate data handling discrepancies in privacy policies for Alexa skills with account linking. Linden et al.~\cite{linden2018privacy} code ICO checklist\footnote{\url{https://ico.org.uk/for-organisations/uk-gdpr-guidance-and-resources/}} into structured queries and leverage the Polisis~\cite{harkous2018polisis} framework to understand the impact of GDPR on privacy policies.

Mori et al.~\cite{9842614} propose a method for using convolutional neural networks (CNNs) to analyze privacy policies and classify them based on compliance with legal requirements. However, the ``black-box" nature of CNNs and the need for adjustments when applying the method to different legal frameworks pose challenges. Rabinia and Nygaard explore utilizing Natural Language Inference (NLI) for compliance checking, demonstrating that models trained on diverse datasets perform better with real-world privacy policy tasks~\cite{Rabinia2022Compliance}.

Existing works have taken a fragmented approach to the compliance problem, focusing on a subset of privacy regulations, not being able to consider the full context of policy texts, and using opaque models that cannot provide reasoning. In contrast, \name addresses these problems by considering a more comprehensive representation of the privacy policy as a truth table of logic formulae. This representation allows for more interpretable compliance with a wider set of privacy regulations.
\section{Logic Representation}

\subsection{Notation}
Let $\mathcal{A}$ be that set of atomic formulae of a logic system, representing its vocabulary. Let $\Sigma$ be the set of logical connectives which can be unary, binary or $n$-ary. Connectives are used to construct formulae, for example, if $\neg$ is an unary connective, for any formula $\phi$, $\neg \phi$ is also a valid formula. Similarly, for a binary connective $\land$, any formula pair $\phi$ and $\psi$ imply that $\phi \land \psi$ is also a valid formula. The set of atomic formulae along with the connectives define $G = (\mathcal{A}, \Sigma)$, the \textit{formulation grammar} of the logic. Let $\Phi_G$ be the set of all possible formulae that can be constructed using the grammar $G$.

\subsection{Valuation Function}
All formulae can be evaluated to either True or False depending on the world model $M$. In other words, if $M$ satisfies a formula $\phi$ i.e., $M \models \phi$, the formula $\phi$ will be evaluated to be True. We define the Base Valuation Function $\texttt{Val}_{M} : \mathcal{A} \rightarrow \{0, 1\}$ as
\[
\texttt{Val}_{M}(\phi) = 
\begin{cases}
    1 & \text{if } M \models \phi \\
    0 & \text{if } o.w.
\end{cases}
\]
where $M$ is the world model and $\phi \in \mathcal{A}$ is an atomic formula.

Next, we extend the valuation function $\texttt{Val}_{M}$ to a function $\texttt{Val}_{M}^* : \Phi_G \rightarrow \{0, 1\}$ that assigns truth values to all formulae generated using grammar $G$. Note that all formulae are built inductively from the set of atomic formulae $\mathcal{A}$. Therefore, the valuation of any complex formula can be determined by recursively applying the logical connectives according to their rules on the valuations of the atomic formulas, i.e.
\[
\texttt{Val}_{M}^*(\oplus(\phi_1,...,\phi_n)) = f_{\oplus}(\texttt{Val}_{M}(\phi_1),...,\texttt{Val}_{M}(\phi_n))
\]

where $\oplus \in \Sigma$ is an $n$-ary connective, $f_{\oplus}: \{0,1\}^n \rightarrow \{0,1\}$ is the truth function associated with the connection $\Sigma$, $\phi_1,...,\phi_n \in \mathcal{A}$ are atomic formulae.

For instance, let $\phi$ and $\psi$ be atomic formulae with valuations $\texttt{Val}_{M}(\phi)$ and $\texttt{Val}_{M}(\psi)$ respectively. For example, the valuations associated with the connectives $\neg, \land$ and $\lor$ can be determined by the following:
\[
\texttt{Val}_M^*(\neg\phi) = 1 - \texttt{Val}_M(\phi)
\]
\[
\texttt{Val}_M^*(\phi \land \psi ) = \text{min}\left(\texttt{Val}_M(\phi), \texttt{Val}_M(\psi)\right)
\]
\[
\texttt{Val}_M^*(\phi \lor \psi ) = \text{max}\left(\texttt{Val}_M(\phi), \texttt{Val}_M(\psi)\right)
\]

\subsection{Logical Representation}
Evaluating whether the world model $M$ satisfies a formula or not, provides information about the model. For instance, if a world model satisfies the formula ``\textit{All entities must have unique identifiers}'', it indicates a structured environment where each entity can be distinctly identified. This suggests that evaluating a large number of such formulae should give a comprehensive representation of the world model.
Therefore, the set of atomic formulae along with the Valuation function -- $(\mathcal{A}, \texttt{Val}_M)$, provides a logical representation of the world model $M$. Any complex formula can then be evaluated by simply combining the atomic valuations using truth functions associated with the connectives of the logic grammar. This representation has the following benefits:

\begin{enumerate}
    \item \textbf{Abstraction:} It provides a concise intermediate representation of the world model. It can act as a proxy for the world model to perform downstream analyses such as Consistency and Compliance (\autoref{sec:logic_def}).
    
    \item \textbf{Efficiency:} Evaluating a formula directly on the world model using the valuation function can be costly -- it requires analysis of the world model. By design, our formulation provides a solution by first making the minimum set of necessary valuations and then using them to evaluate any future formulae. 

    \item \textbf{Explainability:} Valuations for complex formulae can seem opaque and hard to interpret. The atomic formulae are fundamental and can be directly inferred from the world model. Therefore, representing complex formulae as a function of the atomics offers better explanations for formula evaluations.
    
\end{enumerate}

\subsection{Definitions}\label{sec:logic_def}

\noindent \textbf{Consistency.} A world model $M$ is consistent with another world model $M'$ with respect to the logic grammar $G$, denoted as $M \sim_G M'$ if
\[
\forall \phi \in \mathcal{A}, \; \texttt{Val}_M(\phi) = \texttt{Val}_{M'}(\phi)
\]
This means that world models $M$ and $M'$ agree on valuations of all formulae in $\mathcal{A}$, i.e., they have the same logical representation. For example, consider a set of logical statements describing a company's data retention policy, such as ``Retention period for user data is unspecified'', ``Retention period for user data is one year'', and ``Retention period for user is 10 years''. This set of statements can be used to evaluate whether two data retention policies are similar/consistent in terms of how long user data is kept. Similarly, it can also be used to evaluate consistency between different versions of the same data retention policy over time.

\noindent \textbf{Compliance.} Consider a set of formulae $\Phi \subseteq \Phi_{G}$ constructed using the grammar $G$. A world model $M$ is compliant with $\Phi$, denoted as $M \models \Phi$ if
\[ 
\forall \phi \in \Phi, \; M \models \phi
\]

The valuation of each formula in $\Phi$ can be derived by some combination of formulae in $\mathcal{A}$. For instance, the General Data Protection Regulation (GDPR) mandates that personal data should not be kept for longer than necessary for its intended purpose. Now, consider the logical statements from \textit{data-retention} example above. Compliance with the formula ``Retention period for user data is compliant with GDPR'' can be alternatively evaluated by the combination of atomic formulae corresponding to these statements -- ``Retention period for user data is specified'', ``Retention period for user data is no longer than necessary for the purposes it was collected'', and ``User data is securely deleted after retention period expires''. 
\section{PolicyLR: Logic Representation for Privacy Policies}

We present PolicyLR, which is a representation of privacy policies using a set of atomic formulae and a valuation function. In this section, we define PolicyLR's atomic formulae and demonstrate how PolicyLR's representation can be used in downstream applications related to privacy policies.

\subsection{Logical Representation}
To construct PolicyLR's logical representation, we need to define the set of atomic formulae. We leverage existing work on policy annotation that has developed hierarchical taxonomies of privacy concepts~\cite{wilson-etal-2016-creation,arora2022tale}. We specifically use the OPP-115 taxonomy developed by Wilson et al.~\cite{wilson-etal-2016-creation} to generate a list of atomic formulae. The taxonomy is widely used in the literature~\cite{harkous2018polisis, zimmeck2019maps, wagner2023privacy, alabduljabbar2021tldr}, and provides a comprehensive set of privacy practices. We note that the PolicyLR framework is independent of the taxonomy a user might choose.

The OPP-115 taxonomy has two levels -- a top level that defines high-level privacy categories like \lstinline{first-party-collection, data-retention}, and a lower level that defines a set of attributes for each high level category. Each attribute is a categorical variable that can take one of a fixed set of values. For example, the high level category \lstinline{data-retention} has attributes \lstinline{retention-period, retention-purpose} and \lstinline{information-type}. The lower level attribute \lstinline{retention-period} can take one of the values -- $\{\texttt{indefinite},\; \texttt{stated}, \;\texttt{limited}, \;\texttt{unpsecified}\}$.

We next describe how we construct the building blocks of PolicyLR using this taxonomy.

\noindent \textbf{Atomic Formulae.} We define the set of atomic formulas using a space of finite-domain variables. 
We use the high-level privacy categories as the finite-domain variables. For instance, $\texttt{data-retention}(\texttt{period}=\texttt{stated}, \texttt{purpose}=\texttt{advertising}, \texttt{type}=\texttt{location})$ is an atomic formula.

Consider a high-level category $p \in \mathcal{P}$, with lower-level attributes given by $attr(p) \in \mathcal{Q}$. Each attribute $q \in \mathcal{Q}$ can assume a finite set of values $dom(q) \in V$. Now, the set of atomic formulae $\mathcal{A}$ is given by 
\[
\left\{ 
p \left( q_1 = v_1, \ldots, q_n = v_n \right) 
\;\middle|\;
\begin{array}{l}
p \in \mathcal{P}, \{q_1, \ldots, q_n\} = attr(p) \\
v_i \in \text{dom}(q_i) \; \forall i \in \{1, \ldots, n\}
\end{array}
\right\}
\]
where $p$ is a high-level category from $\mathcal{P}$, $attr(p)$ represents the set of attributes associated with $p$, and each $q_i$ is an attribute that can take values from its respective domain $dom(q_i)$. This set $\mathcal{A}$ includes all possible combinations of attributes and their values for each high-level category.

For example, consider a toy taxonomy that only contains \textit{Data Retention} and \textit{First Party Collection} as high-level categories. Furthermore, \textit{Data Retention} only contains two values each for \textit{retention-period} and \textit{retention-purpose} whereas \textit{First Party Collection} consists of \textit{identifiability} and \textit{purpose}.

Our key insight is that given the comprehensiveness of the taxonomy, the set of atomic formulae along with their valuation serves as a complete representation of the privacy policy allowing us to perform a variety of the downstream tasks (\ref{sec:compliance} and \ref{sec:consistency}).

\subsection{Downstream Applications}
Next, we demonstrate how PolicyLR can be useful for downstream applications.

\noindent \textbf{Compliance Analysis.} PolicyLR can help with compliance analysis by mapping the regulatory requirements to logical formulae and allowing to systematically check if the privacy policy adheres to data handling practices and user controls. This can significantly streamline the compliance analysis process and ensure alignment with evolving regulations. 

\noindent \textbf{Consistency Analysis.} PolicyLR can also facilitate consistency analysis within a privacy policy. The logical representation of the policy using atomic formulae enables automated checks for inconsistencies. For instance, the valuation function can reveal if the policy states that a certain type of data is collected for one purpose, but later contradicts itself by allowing the use of that data for a different purpose.

\noindent \textbf{Privacy-based Comparison Shopping.} To perform tasks on a daily basis, customers are often faced with the decision to choose between several applications or services that perform the same task. Apart from quality or cost, these decisions can also be based on the privacy practices of the service provider~\cite{konig2012extending}. \name provides a natural way to compare the privacy policies of multiple services. The atomic formulae can be easily combined to perform comparisons along multiple dimensions.
\section{Compiling Unstructured Documents to PolicyLR}

\begin{figure*}[ht!]
    \centering
    \includegraphics[width=\linewidth]{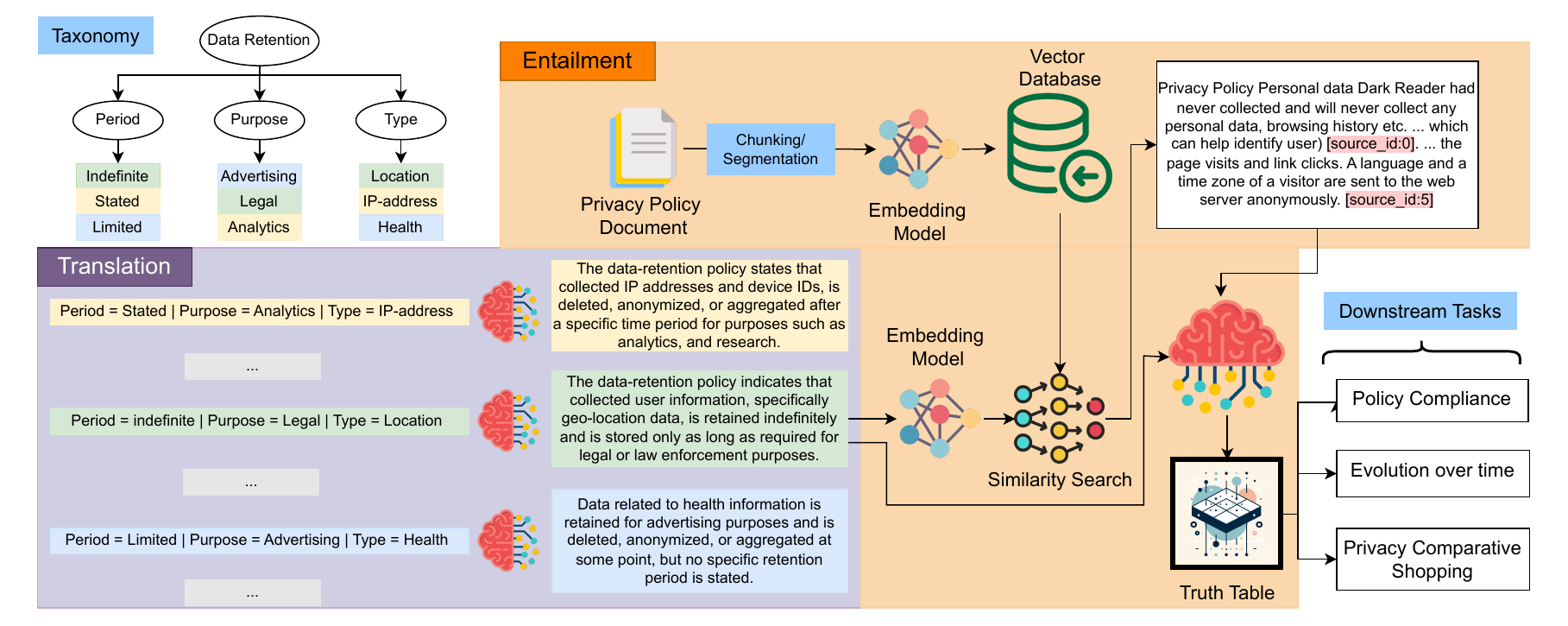}
    \caption{End to end pipeline for \name. We first instantiate \name's atomic formulae using the OPP-115 taxonomy. Each combination of attribute-value pairing becomes an atomic formula. The translation module then transforms each of these into natural language statements. Statements are then compared against the privacy policy text by the entailment module to generate \name's truth table.}
    \label{fig:retrieval}
\end{figure*}

In the previous section, we argued that privacy policy applications can benefit from \name's logical representation. However, existing policies are comprised of long, unstructured text documents that are hard to parse or comprehend by the end users~\cite{mcdonald2008cost,reidenberg2015disagreeable,pollach2007s, linden2018privacy}. In order to make \name relevant to the existing landscape, we need a \textit{compiler} to transform unstructured privacy policies to \name's logical representation. This compiler is essentially a valuation function that can a) handle a privacy policy text document as the world model, b) process the unstructured text, and c) generate valuations for atomic formulae.

Building a valuation function for unstructured text is similar to the Natural Language Inference (NLI) task~\cite{maccartney2009natural}. While NLI involves inferring the logical relationship between two sentences, PolicyLR's compiler needs to do this for a formula and a long policy text document. We reduce the compilation task to an NLI task by leveraging a two-stage pipeline -- (1) \textit{Translation}: We transform both the formula and the policy text document into  sentences suited for the NLI task, and then (2) \textit{Entailment}: We use NLI to ascertain whether the policy text entails the formula. Next, we show how we implement the two stages using Large Language Models. \autoref{fig:retrieval} shows the end to end pipeline for \name along with \textit{translation} and \textit{entailment} modules.

\subsection{Translation}
To reduce compilation to an NLI task, we need to transform the inputs into short sentences that can be used for NLI. For translating a formula, we use the In-Context Learning (ICL) ability of LLMs. For the policy text, we use embedding-based similarity search to extract the relevant sentences from the long policy text. 

\noindent \textbf{Formulas.} Large language Models have remarkable in-context learning abilities~\cite{brown2020language}. It allows LLMs to be applied to new tasks using only a few natural language demonstrations, a phenomenon known as few-shot learning. More concretely, we use a set of $k$ input-output pairs $\{(x^{i}, y^{i})\}_{i=1}^k$, where $x^i$ are arbitrary formulas from PolicyLR's grammar, and $y^i$ are the corresponding natural language translations. We only need a few in-context samples for demonstration, which are crafted by privacy policy experts. For example, the atomic formulae, $\texttt{data-retention}(\texttt{period}=\texttt{indefinite}, \texttt{purpose}=\texttt{legal}, \texttt{type}=\texttt{location}$ is translated into ``\textit{The data-retention policy indicates that
collected user information, specifically
geo-location data, is retained indefinitely
and is stored only as long as required for
legal or law enforcement purposes.}''. Note that translation of formulas is independent of the privacy policy texts and therefore, only needs to be done once.

\noindent \textbf{Policy Text.} While recent LLMs have long context windows, privacy policies might still be too long to fit within the LLM's context window. Therefore, to perform NLI using LLMs, we need to retrieve only the excerpts of the privacy policy relevant to the formula being entailed. We do this via a Retrieval-Augmented Generation (RAG) methodology. We first chunk the long privacy policy text into shorter segments. Then, we approximate the semantics using text-embedding models. This segment embedding mapping is stored in a vector database, to allow for quick retrieval during the entailment process. During retrieval, we ensure that additional context is not lost due to chunking by adding the previous and the next segment of the retrieved segment. Note that this segmentation and embedding process only needs to be done once per privacy policy, subsequently allowing for entailment of multiple formulas.

\subsection{Entailment}
Once we have the translations for the formula and the privacy policy, we can use NLI to infer the formula's valuation with respect to the policy. Specifically, we prompt an LLM with the following, \textit{``According to the Privacy Policy $P$,  is the following statement True? $Q$''}. Here, $P$ and $Q$ are translations of the policy text and formula respectively. To extract $P$, we first compute the embedding vector of $Q$ and then fetch the most similar $k$ segments from the segment-embedding mapping computed in the translation stage. We append the tag ``$\text{[source\_id:i]}$'' at the end of the $\text{i}^{th}$ segment and concatenate them to get the condensed and most relevant policy text $P$. We then augment the prompt to provide evidence by highlighting the segments that were used by the LLM to perform the entailment task, using the template -- ``\textit{Give evidence by providing all the source ids that are used to answer the question in the format of - Evidence:[2,3,7,...]}''. This evidence makes the valuation more interpretable and can be useful for downstream applications. Note that $k$ here is a tunable parameter that controls the context ($P$) provided to the LLM. It can be tuned to balance the precision-recall trade-off for any downstream task. For example, increasing the value of $k$ will result in higher recall but can lower the precision. We discuss this further in~\autoref{sec:tosdr_entail}.
\section{Evaluation}

\subsection{Research Questions}
\begin{enumerate}
\item[\textbf{Q1.}] \textbf{What is the performance of \name's compiler?}\\
We demonstrate that \name is able to successfully perform valuations against unstructured privacy policy text documents. When evaluated on $2656$ entailment instances from ToS;DR, a privacy community annotated dataset, using an open-source LLM gemma2-27b, \name's compiler achieves a precision value of $0.84$ and recall value of $0.88$. 
\smallskip
\item[\textbf{Q2.}] \textbf{How effective is \name for the Policy Compliance task?} \\Next, we show the effectiveness of \name on policy compliance tasks using two existing datasets. Specifically, we analyze the compliance of $419$ privacy policies with respect to $3$ compliance rules based on the Article 13 of the GDPR. We find that \name achieves an average F-1 score of $0.91$ across the two datasets, highlighting the efficacy of \name in downstream tasks.
\smallskip
\item[\textbf{Q3.}] \textbf{How can \name be used to perform inconsistency detection and privacy comparison shopping?}\\
We analyze privacy policies of the popular apps on the Google Play Store to show how \name can be used to a) Analyze the consistency of privacy policies over time and b) Compare privacy practices of apps in similar categories. We note that the latter can be used to provide a comparative view of the privacy of apps that can act as a signal for users when deciding which apps to buy or use.

\end{enumerate}

\subsection{Experimental Setup: Datasets}
We use the following experimental setup for answering each of the research questions using the datasets described in \autoref{fig:dataset_table}.

\begin{table}[t]
\footnotesize
\begin{center}
\begin{tabular}{l|c|c}
\toprule
\textbf{Task} & \textbf{Dataset} & \textbf{$\#$ Unique Policies} \\
\midrule
Compiler Entailment & ToS;DR Annotations & 1074 \\
\midrule
\multirow{2}{0.9in}{Policy Compliance}  & OPP-115 & 115 \\
& AutoCompliance & 304 \\
\midrule
Consistency Analysis & Self Curated & 85 \\
\bottomrule
\end{tabular}
\end{center}
\caption{The analyzed datasets in the evaluation.}
\label{fig:dataset_table}
\end{table}

\noindent{\textbf{ToS;DR Dataset.}}
PolicyLR's compiler provides a valuation function that can evaluate formulae against an unstructured privacy policy text document. To evaluate its performance, we use the ToS;DR dataset. Terms of Service; Didn't Read (ToS;DR) is a collaborative, community-driven platform where users and volunteers contribute to evaluating and summarizing terms of service and privacy policies~\cite{roy2012terms}. The platform helps make privacy policies easier to understand by using crowd-sourced annotations. Users sign up on the platform to annotate policies by linking parts of the policy to specific data practices called \textit{cases}. \textit{Cases} are concise statements about privacy settings, for example - `\textit{You can delete your content from this service}' or `\textit{This service tracks you on other websites}'. A moderator then reviews these matches and either approves them or provides feedback. We can use \textit{cases} as proxies for the natural language translations of logical formulae. The moderator-approved matches provide reliable ground truth for their valuation against privacy policy texts.

The dataset consists of $246$ cases, which are used to annotate $1074$ unique privacy policy texts. After approval from the moderators, this leads to a total of $13179$ case and privacy policy pairings. ToS;DR comprises only positive instances. To construct negative instances (i.e., where the case doesn't match the policy text), we manually analyze all the cases and find $11$ pairs of mutually contrasting cases. This means that if a case is evaluated to be true for a policy text, its contrasting case will necessarily be evaluated to be false on that text. Using this formulation, we construct a total of $1222$ negative instances. We sample a random subset of size $1300$ from the positive instances to get a total of $1522$ case-policy pairs. For each instance, we apply PolicyLR's compiler to evaluate the case against the policy text, answering either true or false. 

\subsection{Experimental Setup: Implementation Details of PolicyLR}
We now describe how PolicyLR is instantiated for our experiments.

\noindent{\textbf{Models.}} While evaluating PolicyLR's compiler on the ToS;DR dataset, we consider several open-source LLMs -- 8  billion and 70 billion versions of Meta's Llama3 as well as 9 billion and 27 billion versions of Google's Gemma2. We also evaluate OpenAI's latest closed-source model Gpt4-O. For the Compliance task,  we consider the larger versions of both Llama3 and Gemma2. Finally, for the consistency task, we consider Llama3\footnote{Gemma2 was released very recently (06-27-2024). Due to time constraints, it is only part of our ToS;DR evaluation}. To get deterministic results, we set the sampling temperature to 0 for all models.

\noindent{\textbf{Translation.}}
Since the translation of atomic formulae needs to be done only once, we use the most performant LLM gpt4-o for this task. Below we show the prompt used for this task for one of the \lstinline{data-retention} atomic:

\begin{tcolorbox}
\footnotesize
SYSTEM:

\normalsize
You are a privacy policy expert. A privacy setting consists of a combination of attributes. Each of these has an associated value, along with a description of what that value means. You have to construct a concise statement that describes the setting. Only output the statement.
\footnotesize

USER:
\normalsize

Attribute: \lstinline{period}, Value: \lstinline{limited}, Description: ...

Attribute: \lstinline{purpose}, Value: \lstinline{ads}, Description: ...
\end{tcolorbox}

Here, we provide a description for each attribute value as described in the OPP-115 dataset. 

\noindent{\textbf{Entailment.}} To make it more accessible, we implement \name's entailment module using only open source components. We tune the hyperparameters using a disjoint set of 10 policy documents. We use the \textit{SentenceSplitter} API from LlamaIndex\footnote{https://docs.llamaindex.ai/} to segment the privacy policy text. Each segment comprises $300$ tokens. Then, we generate text embeddings using \lstinline{UAE-Large-V1}~\cite{li2023angle}, which is a popular open-source embedding model. We store these embeddings in \textit{chroma}\footnote{https://www.trychroma.com/} which is a open-source vector database. For any entailment task, we query the database for the top-$k$ segments that are most similar to the embedding of the hypothesis. We use $k=5$ and $k=10$ for evaluating the compiler, and $k=10$ for the rest of the evaluation.

\subsection{Performance of \name's compiler}\label{sec:tosdr_entail}
\noindent \textbf{Entailment.}
We evaluate our compiler on the $1300$ positive and $1222$ negative case-policy pairing from ToS;DR. We parse the response of the entailment LLM and assign it a value of true if it begins with ``\textit{Yes}'' and false if it begins with ``\textit{No}''. We did not observe any instance when response did not begin with either ``\textit{Yes}'' or ``\textit{No}''. This again demonstrates the high instruction following capability of the latest LLMs. \autoref{fig:entail_table} shows the precision, recall and F1 score when using 2 variants of both llama3 and gemma2 as well as the closed source model gpt4-0. For each setting, we show results when providing the LLM with the 5 and 10 most relevant segments from the privacy policy. First, we observe the number of policy segments provides a trade-off between precision and recall. Fewer segments provide a higher precision whereas adding more segments improves recall. This is likely because LLMs struggle with longer contexts~\cite{li2024long}, but also might miss out on relevant context in case of fewer segments. Second, we observe that larger models perform better in the case of both llama3 and gemma2. In terms of F1 score, we find that gemma2 outperforms llama3 and even gpt4-o. Overall, gpt4-o has the best performance with a precision-recall of $0.94$ and $0.84$ respectively. Among open-source models, Llama3-70b has the highest precision. To demonstrate \name's applications, we use Llama3-70b.

\noindent \textbf{Error Analysis.} We perform a deep dive into the errors for gpt4-o to better understand \name's performance. The LLM wrongly entailed a total of $280$ case-policy pairs -- $204$ positive pairs and $76$ negative pairs. There are three primary reasons for these errors -- (1) Insufficient Context, (2) LLM Reasoning Error, and (3) Incorrect Annotation Error. 

\noindent \textit{Insufficient Context Error.} To characterize the first error type, we use the policy excerpt that was used by the ToS;DR moderators to approve the case-policy pairing. The first type of error occurs in instances where the policy excerpt is not part of the context retrieved by the embedding model. In these cases, the LLM does not have the necessary context to correctly entail the case. Note that since we only have policy excerpts for positive pairs, this methodology can only discover insufficient context errors for the positive pairs. Out of the $204$ positive pair errors, $95$ were due to insufficient context. For the other two error types, we perform a qualitative analysis and highlight some interesting cases below.

\noindent \textit{LLM Reasoning Error.} The following belongs to the second error type where the LLM performs incorrect entailment. In this case, the LLM reasoning seems too nit-picky and suggests a lack of common sense. 

\begin{tcolorbox}[colback=yellow!10!white, colframe=red!50!black, title=LLM Reasoning Error, fonttitle=\small, fontupper=\small, boxsep=1mm, left=1mm, right=1mm, top=1mm, bottom=1mm]
    \underline{Case:} This service is only available to users over a certain age.\\
    \underline{Response:} \textit{No. ... it does mention that access may be terminated without warning if it is believed that a user is underage or otherwise ineligible. This implies a concern about age but does not set a specific age limit. ...}
\end{tcolorbox}

\noindent \textit{Incorrect Annotation Error.} The below instance belongs to the third error type where the TsD;DR annotation is wrong. Upon manually analyzing the privacy policy, we find that the policy does not state that anonymization is done for all cookie types. 

\begin{tcolorbox}[colback=yellow!10!white, colframe=red!50!black, title=Incorrect Annotation Error, fonttitle=\small, fontupper=\small, boxsep=1mm, left=1mm, right=1mm, top=1mm, bottom=1mm]
    \underline{Case:} The cookies used by this service do not contain information that would personally identify you.\\
    \underline{Response:} \textit{No. The privacy policy does not ... Anonymous Performance Cookies is aggregated and therefore anonymous, it does not provide a similar assurance for other types of cookies, such as Strictly Necessary Cookies and Security Cookies. ...}
\end{tcolorbox}

\noindent \textbf{Evidence.} We also evaluate the effectiveness of \name's evidence functionality. We again leverage ToS;DR's policy excerpts and evaluate whether the context segments cited in the LLM response contain the excerpt. Out of the $1300$ positive pairs, we found $1056$ instances where the excerpt was part of the retrieved context. We use these $1056$ case-policy pairs to evaluate the evidence functionality. We observe that the LLM, on average, cites $2$ context segments (out of $10$) while responding to each of the above pairs. Overall, we found the excerpt as part of the evidence in $854$ cases, giving a recall of $81\%$.

\begin{table}[t]
\footnotesize
\begin{center}
\begin{tabular}{l|c|c|c|c}
\toprule
\textbf{Model} & \textbf{Top k} & \textbf{Precision} & \textbf{Recall} & \textbf{F1} \\
\midrule
\multirow{2}{0.5in}{llama3-8b}  & 5 & 0.86 & 0.76 & 0.81 \\
& 10 & 0.81 & 0.91 & 0.86 \\
\midrule
\multirow{2}{0.6in}{llama3-70b} & 5 & 0.94 & 0.75 & 0.84 \\
& 10 & 0.94 & 0.81 & 0.87 \\
\midrule
\multirow{2}{0.6in}{gemma2-9b} & 5 & 0.93 & 0.77 & 0.84 \\
& 10 & 0.91 & 0.84 & 0.87 \\
\midrule
\multirow{2}{0.6in}{gemma2-27b} & 5 & 0.92 & 0.83 & 0.87 \\
& 10 & 0.91 & 0.88 & 0.90 \\
\midrule
\multirow{2}{0.5in}{gpt4-o} & 5 & 0.93 & 0.76 & 0.84 \\
& 10 & 0.94 & 0.84 & 0.89 \\
\bottomrule
\end{tabular}
\end{center}
\caption{Performance of the entailment task on ToS;DR data.}
\label{fig:entail_table}
\end{table}

\section{\name for Policy Compliance}
\label{sec:compliance}
Validating the compliance of privacy policies against regulations like GDPR is a critical issue because it ensures the protection of individuals' personal data and maintains their privacy rights. Privacy regulations set stringent standards for data handling, requiring organizations to be transparent about data collection, usage, and storage practices. Non-compliance can lead to significant legal penalties and damage to an organization's reputation. Moreover, ensuring compliance fosters trust between consumers and businesses, as individuals are more likely to engage with companies that respect and protect their privacy. Effective compliance validation also helps organizations avoid data breaches and misuse, thereby safeguarding sensitive information and enhancing overall data security.

We evaluate \name's performance on policy compliance tasks using two annotated privacy policy datasets. Liu et al. ~\cite{liu2021have} extract compliance rules from GDPR Article 13. Analyzing our OPP-115 taxonomy, we find that $5$ out of the $9$ rules can be mapped on the taxonomy. Out of these rules, we discard \texttt{Collect Personal Data $\rightarrow$ Data Processing Purpose} as there are very few instances for this in the ground truth. For both datasets, we evaluate policies on the compliance rules, evaluating a total of $419$ privacy policies. For each compliance rule, we first represent it as a composition of \name's atomic formulae. We note that for some rules, there can be multiple valid compositions. This is because some rules in natural language can be vague and have multiple interpretations. Rules formed using the atomic formulae, on the other hand, are precisely defined by the logic system. \name, then evaluates each formula corresponding to each compliance rule using the valuations of \name's atomic formulae.

\subsection{Compliance Dataset}
\label{sec:compliance_dataset}
 To evaluate PolicyLR's performance on the policy compliance task, we use two existing datasets -- (1) Online Privacy Policies (OPP-115) dataset~\cite{wilson-etal-2016-creation} comprising 115 policies, and (2) AutoCompliance~\cite{liu2021have} corpus comprising 304 policies. The OPP-115 dataset provides 23K sentence-level annotations based on the OPP-115 taxonomy. The annotations are at two levels: the first level consists of paragraph-sized segments annotated as per the high-level categories of the taxonomy. The second level includes parts of segments annotated for attribute-value pairs such as \texttt{retention-period: limited, purpose: advertising,} etc.

 AutoCompliance~\cite{liu2021have}, on the other hand, focuses on policy compliance. They analyze Article 13 of the GDPR and manually extract $10$ labels that discuss personal information collection. They further annotate 304 policies by segmenting the policies and annotating each line of the policy as either one of the $10$ labels or \textit{others}. They then build $9$ compliance analysis rules that measure compliance of a privacy policy with the GDPR. For example, one of the rules is \textit{Collect Personal Information $\rightarrow$ Data Retention Period}, implying that if a policy collects personally identifiable information, then the data retention period must be specified. Finally, to generate the compliance dataset, they obtain policy-level annotation by aggregating the segment-level annotation.
 
 Note that some of the rules can be decomposed into combinations of low-level categories of the OPP-115 taxonomy. For instance, "\textit{Collect Personal Info $\rightarrow$ Contact Details}" can be represented in OPP-115 taxonomy as: \texttt{first-party-collection - identifiability: identifiable} AND \texttt{data-retention - retention-period: not unspecified}. We note that while both these datasets provide sentence-level annotations, we follow the aggregation-based formulation similar to Liu et al. ~\cite{liu2021have} Linden et al. ~\cite{linden2018privacy} to get policy-level annotations. For instance, the data retention period annotation of the entire privacy policy can be derived using the presence of at least one sentence, which is annotated for the data retention period. We also note that while evaluating \name, we provide the entire privacy policy and use the aggregated label as the ground truth.

\begin{table*}[t]
\footnotesize
\begin{center}
\begin{tabular}{lccccc}
\toprule
\textbf{Compliance Rule} & \textbf{Regulation} & \textbf{Formula} & \textbf{Precision} & \textbf{Recall} & \textbf{F1} \\
\midrule
\begin{tabular}[l]{@{}l@{}}\textit{Contact details of the data} \\ \textit{controller should be provided}\end{tabular}
& $\texttt{GDPR Art 13.2(a)}$ & $\texttt{rp} = \texttt{stated} \lor \texttt{rp} = \texttt{limited} \lor \texttt{rp} = \texttt{indefinitely}$ & 0.83 & 0.94 & 0.88 \\
\midrule
\begin{tabular}[l]{@{}l@{}}\textit{Contact details of the data} \\ \textit{controller should be provided}\end{tabular} & $\texttt{GDPR Art 13.1(a)(b)}$ & $\texttt{contact-information} = \texttt{present}$ & 0.90 & 0.94 & 0.92 \\
\midrule
\multirow{2}{1.2in}{\textit{Users should be able to modify/delete their data}}  & \multirow{2}{*}{\lstinline{GDPR Art 13.2(b)}} & $\texttt{acc} = \texttt{edit} \lor \texttt{acc} = \texttt{deactivate} \lor \texttt{acc} = \texttt{delete}$ & 0.86 & 0.98 & 0.91 \\
& & $\texttt{acc} = \texttt{edit} \land \texttt{acc} = \texttt{deactivate} \land \texttt{acc} = \texttt{delete}$ & 0.94 & 0.42 & 0.58 \\
\bottomrule
\end{tabular}
\end{center}
\caption{Compliance Task}
\label{table:compliance_table}
\end{table*}

\subsection{Compliance Results}
Table~\ref{table:compliance_table} shows the compliance rule, the corresponding regulation, the computed formula as a composition of \name's atomic formulae, and the performance of \name for each rule. We find that \name has an average F1 score of $0.90$ with an average recall of $0.95$, outperforming Autocompliance by $4\%$. We note that \name uses off-the-shelf open-source LLM as opposed to Autocompliance, which trains custom classifiers for each of the rules. Further, we observe that in Autocompliance, training data for the classifiers is also used for compliance analysis, whereas in \name, we do not perform any training and use the entire dataset for evaluation. 

We also analyze \name's performance on the \textit{Right to Access compliance rule} (\texttt{Collect Personal Information $\rightarrow$ Right to Access}), achieving an F1-score of only $0.51$. Upon closer examination of the dataset, we find that several annotations for this class appear to be incorrect. This discrepancy might be attributed to the inherent similarity between the Right to Access and the Right to Delete, as a similar trend is observed within the OPP-115 dataset. Consequently, we opt to exclude the \textit{Right to Access rule} from this analysis for the sake of clarity.

\begin{table*}[t]
\footnotesize
\begin{center}
\begin{tabular}{lccccc}
\toprule
\textbf{Compliance Rule} & \textbf{Regulation} & \textbf{Formula} & \textbf{Precision} & \textbf{Recall} & \textbf{F1} \\
\midrule
\begin{tabular}[l]{@{}l@{}}\textit{Data retention period} \\ \textit{should be specified}\end{tabular}
& $\texttt{GDPR Art 13.2(a)}$ & $\texttt{rp} = \texttt{stated} \lor \texttt{rp} = \texttt{limited} \lor \texttt{rp} = \texttt{indefinitely}$ & 0.88 & 0.73 & 0.80 \\
\midrule
\begin{tabular}[l]{@{}l@{}}\textit{Contact details of the data} \\ \textit{controller should be provided}\end{tabular} & $\texttt{GDPR Art 13.1(a)(b)}$ & $\texttt{contact-information} = \texttt{present}$ & 0.98 & 0.95 & 0.97 \\
\midrule
\multirow{2}{1.2in}{\textit{Users should be able to modify/delete their data}}  & \multirow{2}{*}{\lstinline{GDPR Art 13.2(b)}} & $\texttt{acc} = \texttt{edit} \lor \texttt{acc} = \texttt{deactivate} \lor \texttt{acc} = \texttt{delete}$ & 0.88 & 0.95 & 0.92\\
& & $\texttt{acc} = \texttt{edit} \land \texttt{acc} = \texttt{deactivate} \land \texttt{acc} = \texttt{delete}$ & 1.00 & 0.41 & 0.58 \\
\bottomrule
\end{tabular}
\end{center}
\caption{Compliance Task on privacy policies from OPP-115 dataset.}
\label{table:compliance_opp}
\end{table*}

Next, we present \name's performance on the OPP-115 compliance dataset in Table~\ref{table:compliance_opp}. We find that \name achieves an average recall of $0.88$ and an average F1-score of $0.92$. It is worth noting that \name's performance on the data retention compliance rule falls below the overall average. This can be attributed to the inherent limitations within the OPP-115 dataset, where data quality for data retention information is known to be low. Prior research has documented similar challenges; for instance, Polisis~\cite{harkous2018polisis} reports an F1-score of only 0.71 for the same rule using a classifier trained on OPP-115. We manually analyzed the errors and found that the majority of them were due to incorrect annotation. For example, the segment: \textit{Email share feature may be saved for your convenience for future articles you may wish to email} was tagged as \texttt{retention-period: limited}. 

\noindent\textbf{Error Analysis.} Investigating the errors in the compliance task for both datasets, we find that errors can be categorized into two classes below.

\noindent \textit{LLM Reasoning Error.} In these errors, we observe that the LLM extrapolates in the reasoning and reaches the wrong conclusions. Such instances are not common and generally indicate incorrect assumptions by the LLM. For example, consider the following task for \lstinline{alibaba.com}: 

\begin{tcolorbox}[colback=yellow!10!white, colframe=red!50!black, title=LLM Reasoning Error, fonttitle=\small, fontupper=\small, boxsep=1mm, left=1mm, right=1mm, top=1mm, bottom=1mm]
    \underline{Hypothesis:} Users can delete their account, but the company/organization may continue to keep some of the user's data.\\
    \underline{Response:} \textit{Yes. ... The privacy policy states that We retain your Personal Data as long as we have an ongoing legitimate business need to do so... When we have no ongoing legitimate business need to process your Personal Data, we will either delete or anonymise it... This implies that users can request to delete their account... company may continue to keep some of user's data ...}
\end{tcolorbox}
Here, the policy never mentions that users can request to delete the data, but the LLM infers that because the company may delete the data on their own. We note that these errors can be potentially addressed by choosing a more capable LLM, such as GPT4 or Gemma2.

\noindent \textit{Annotation Error.} These are the errors where the human annotators made a mistake and mislabeled a segment, but the LLM correctly categorizes it. For example:

\begin{tcolorbox}[colback=yellow!10!white, colframe=red!50!black, title=Annotation Error, fonttitle=\small, fontupper=\small, boxsep=1mm, left=1mm, right=1mm, top=1mm, bottom=1mm]
    \underline{Case:} Data retention is limited, meaning data is deleted, anonymized, or aggregated at some point, but no specific retention period is stated.\\
    \underline{Response:} \textit{Yes. ... policy does mention that data will be deleted or removed in certain circumstances. For example, it states that When your account is deleted, your user profile and all public activities (comments, conversation topics) will be removed. ...}
\end{tcolorbox}
Here, the LLM correctly infers that since the account data is deleted when the account is deleted, the retention period is \textit{limited}. 

Note here that we used the open source LLM \textit{Llama-70b} to perform compliance analyses in this section. The performance of closed-source models, such as GPT-4 and Gemini, is believed to be better than open-source models. This suggests that performance can be further improved by using a better model. 

\noindent\textbf{Precision-Recall Tradeoff.}
We acknowledge the inherent ambiguity present in natural language queries. This ambiguity can lead to multiple valid formulas representing a single compliance rule, as shown in \autoref{table:compliance_table}. The rule, "\textit{Users should be able to modify/delete their data,}" can be translated into two distinct formulas.
\[
\texttt{acc} = \texttt{edit} \lor \texttt{acc} = \texttt{deactivate} \lor \texttt{acc} = \texttt{delete}
\]
\[
\texttt{acc} = \texttt{edit} \land \texttt{acc} = \texttt{deactivate} \land \texttt{acc} = \texttt{delete}\]
These formulas offer a mechanism for controlling the desired level of precision in our analysis. For instance, the first formula allows for any combination of edit access, delete access, or account deactivation to satisfy the rule. Conversely, the second formula implements a stricter interpretation, requiring all three functionalities to be present. This interplay between formula composition and performance reflects the precision-recall tradeoff - as the level of restrictiveness (precision) increases, the ability to identify compliant policies (recall) decreases. 
By allowing the creation of multiple formulas for a single rule, the user can tailor the analysis to specific requirements.

\begin{figure}[ht!]
    \centering
    \includegraphics[width=\linewidth]{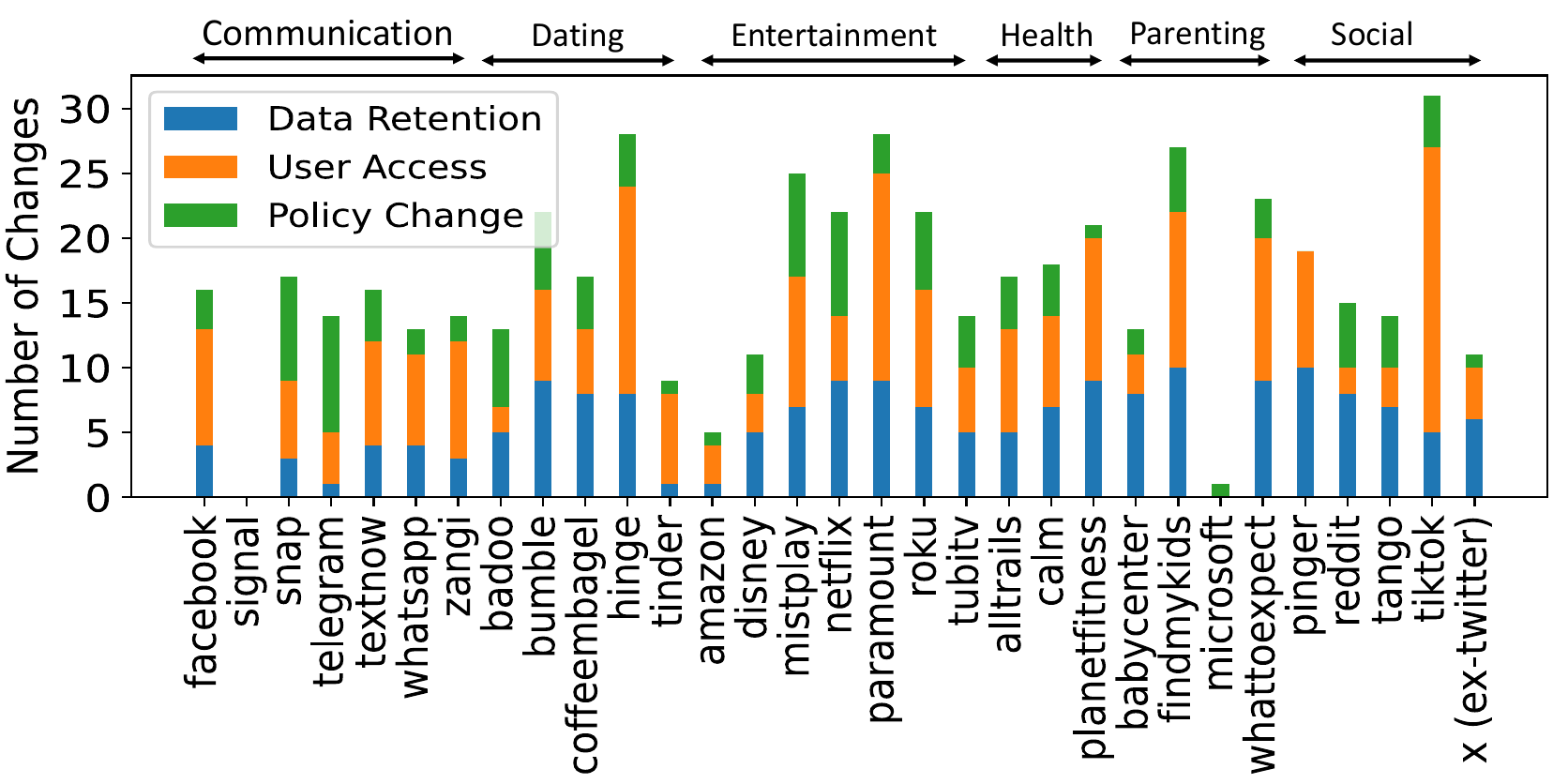}
    \caption{Number of atomic formulae where the valuation is different between historical and current version of privacy policies of 31 Google Play Apps. The large number of changes are due to the introduction of the GDPR regulations. \name provides a way to perform a fine-grained analysis of the evolution of privacy policies in response to new regulations.}
    \label{fig:time_inc}
\end{figure}

We note that prior works in compliance analysis create structured rules~\cite{linden2018privacy} or train individual classifiers~\cite{liu2021have}.  However, these approaches perform classification at a segment level and do not consider the full context of policy texts. \name addresses these limitations and leverages the truth table to efficiently perform the compliance analysis. Additionally, the existing approaches only cater to a subset of privacy regulations. \name allows a more comprehensive analysis by leveraging the joint distribution of privacy practices.

\section{\name for Policy Consistency}
\label{sec:consistency}

Evaluating the consistency of privacy policies is an important and well-studied problem~\cite{linden2018privacy}. It involves comparing the practices of two policy documents, which could be different versions of the same privacy policy or belong to different applications. Consistency among different policy texts of the same app is essential to ensure clarity in privacy practices. Comparing different historical versions of the same policy text can help in understanding the evolution of practices over time. Finally, comparing policies from different apps can help enable privacy-based comparative shopping~\cite{konig2012extending}, where users can pick and choose apps based on their privacy preferences. 

Prior works~\cite{linden2018privacy, wagner2023privacy} have studied the evolution of privacy policy by creating structured rules for coverage, specificity and compliance, and comparing the practices along these dimensions before- and after- GDPR. However, their main limitation stems from using sentence-level natural language processing techniques, which are not capable of performing joint classification on complex aspects like \textit{data retention period for location data used for advertising purposes}. Prior work on comparing practices across documents also suffers from similar limitations~\cite{konig2012extending,khandelwal2023overview}.

\name's logical representation provides a formal way to perform consistency analysis (\autoref{sec:consistency}). Using \name, we can directly compare the atomic formulae valuations of two different versions of a policy document. \name overcomes the limitations of prior work by incorporating a joint distribution of privacy attributes, which are automatically integrated from existing policy taxonomies. Below, we describe two applications of the consistency formulation of \name: a) Evolution of Privacy Practices Over time, and b) Privacy Comparative Shopping.

\subsection{Policy Dataset for Consistency Analysis}
\label{sec:consistency_dataset}
 To demonstrate how PolicyLR can assist with the consistency analysis of privacy policies, we curated a dataset of popular privacy policy texts from \lstinline{play.google.com}. For differential analysis, we first manually collected privacy policy URLs for the top 10 applications from \lstinline{play.google.com} for each of the following categories -- \textit{Communication}, \textit{Dating}, \textit{Entertainment}, \textit{Health}, \textit{Parenting} and \textit{Social}. We found a total of $54$ unique policy URLs among all these categories. Note that different apps from the same parent company can have an identical privacy policy, resulting in fewer unique policies. For example, \textit{Instagram} and \textit{Meta} have identical privacy policies. 

To curate a dataset for the evolution of privacy practices over time, we use the Wayback Machine\footnote{\url{https://web.archive.org/}} from the Internet Archive to get the historical versions of the privacy policies for the apps above. For an earlier timestamp, we purposefully choose the latest policy before 2018 to obtain a pre-GDPR version of the policy. As described by Linden et al.~\cite{linden2018privacy}, the majority of the updates occurred in 2018. Selecting a pre-GDPR version of the policy allows us to observe the effect of the GDPR. Interestingly, due to website migrations, we find that several websites' current privacy policy URLs did not exist back in 2017. To get the old policies for these cases, we manually crawled the homepage using Wayback Machine and identified the privacy policies. Following this approach, we found a valid historical version for $31$ out of $54$ URLs. We downloaded the raw HTML for each of the collected URLs and used \textit{Beautiful Soup}\footnote{\url{https://www.crummy.com/software/BeautifulSoup/}} library to extract the associated text. This resulted in a total of $85$ unique privacy policy texts.

Next, we discuss the two downstream applications of the consistency framework of \name.

\subsection{Evolution of Privacy Practices Over Time}
We demonstrate using \name to analyze the evolution of privacy practices of mobile apps over time. This is an important analysis because by tracking changes in disclosure of privacy practices, we can get insights into how companies adapt to evolving legal landscapes. This analysis can also assist policy auditors in the enforcement of regulations to ensure effective user privacy protection. Additionally, it can provide the user with a better understanding of how their data is handled by the apps they utilize.

We analyze the \textit{Policy Dataset for Consistency Analysis}~\ref{sec:consistency_dataset} for this task. Specifically, we analyze pre-GDPR and post-GDPR policies of the $31$ apps in the dataset. The dataset consists of 31 top applications on the Play Store. We focus on three high-level categories from the OPP-115 taxonomy, namely, \texttt{Data Retention}, \texttt{User Access Edit and Deletion}, and \texttt{Policy Change}. Recall that high-level categories in OPP-115 are independent of each other, and therefore, any combination of these high-level categories is a valid set of atomics for \name. We choose to restrict ourselves to only these three categories because the number of atomic formulae can grow exponentially with the number of leaf nodes.

\noindent\textbf{Results.} We implement \name using the open source \textit{Llama-70B} model as the compiler with the custom taxonomy defined above. We then take the two valuation functions and compare the valuation of each atomic formula to identify the inconsistencies. \autoref{fig:time_inc} shows the distribution of changes observed in privacy practices across the three high-level categories for each app. Notably, these categories exhibit some overlap with the information mandated by GDPR Article 13. We find that policies across all app categories have changed over time. For instance, the "\texttt{User Access}" category witnessed the most significant change in the case of TikTok, followed by Hinge and FindMyKids. This indicates that these applications potentially made substantial adjustments to user access controls in response to the regulation. We confirmed this by manually checking the policies. Conversely, Microsoft's privacy policies exhibited minimal changes between the pre- and post-GDPR versions. Similarly, Amazon's policies remained largely unchanged in the "\texttt{Data Retention}" category, while some adjustments were made in the "\texttt{User Access}" category. 

\noindent\textbf{Examples of Changes in Practices.} We manually select the following two apps: \textit{Facebook} and  \textit{Tiktok} and show examples of policy change. 
For example - in 2017, \textit{Facebook's} policy mentioned that they can retain information indefinitely for security reasons. However, the latest policy mentions that the data can be deleted upon request. 

\begin{tcolorbox}[colback=yellow!10!white, colframe=red!50!black, title=Facebook Messanger: Change in Data Retention Period , fonttitle=\small, fontupper=\small, boxsep=1mm, left=1mm, right=1mm, top=1mm, bottom=1mm]
    \underline{Pre-GDPR:} ...We may also access, preserve and share information when we have a good faith belief it is necessary to: detect, prevent and address fraud and other illegal activity...\\
    \underline{Post-GDPR:} \textit{We keep information for as long as we need it to provide a feature or service. But you can request that we delete your information. We'll delete that information unless we have to keep it for something else, like for legal reasons.
}
\end{tcolorbox}

Similarly, \textit{Tiktok's} policy in 2017 did not have any means by which user could delete their data, whereas in 2024, they include a full section on \textit{Your Rights}. 

\begin{tcolorbox}[colback=yellow!10!white, colframe=red!50!black, title=Facebook Messanger: Change in Data Retention Period , fonttitle=\small, fontupper=\small, boxsep=1mm, left=1mm, right=1mm, top=1mm, bottom=1mm]
    \underline{Pre-GDPR:} ...No mention of user rights...\\
    \underline{Post-GDPR:} \textit{You may submit a request to know, access, correct or delete the information we have collected from or about you...You may also exercise your rights to know, access, correct, delete, or appeal by sending your request to the physical address...
}
\end{tcolorbox}

We presented here a proof-of-concept analysis that showcased the versatility of \name by analyzing the evolution of privacy practices over time. We acknowledge the small size of the Policy Dataset as a limitation and leave the full-scale measurement analysis for future work. Our intention here was to perform a case study showing that \name can be used to comprehensively analyze the evolution of privacy practices over time. For example, no other framework is currently capable of identifying the joint distribution of privacy practices in a policy. 

\subsection{Privacy Comparative Shopping}
We now demonstrate how \name can be used as the basis for a privacy-centric shopping assistant by allowing users to compare apps based on their privacy preferences. Users can select specific privacy dimensions such as data retention, data sharing, and encryption practices, and the system will group apps accordingly, presenting a clear comparison of their privacy policies. For example, consider \textit{Alice}, a privacy-conscious person, is considering using a messaging app and is confused between \textit{Signal} and \textit{WhatsApp} as they offer similar functionalities. Before making a decision, \textit{Bob} may want to compare the \textit{Data Retention} practices of the two apps. \name can perform this analysis and provide this information.

We show a proof-of-concept version of this application using the most recent privacy policies from the \textit{Privacy Policy} dataset. This involves performing a holistic comparison of practices across apps. \name's consistency formulation provides a natural way to compare privacy policies by directly comparing \name's atomic formulae valuations. \autoref{fig:privacy_shop} compares the \lstinline{data-retention} practices of $8$ most popular Google Play apps from the Communication category -- \textit{Whatsapp, Snapchat, Messenger, Telegram, Signal, Discord, Textnow} and \textit{Zangi}. We compare apps across two attributes of \lstinline{data-retention}: \lstinline{purpose} and \lstinline{retention-period}. We place an app's icon in a cell if the valuation of the corresponding formula is true. 

\begin{table}[t]
\footnotesize
\begin{center}

\begin{tabular}{l?c|c|c|c}
$\texttt{purpose}$ & \multicolumn{4}{c}{$\texttt{retention-period}$}
\\
\cline{2-5}
 & $\texttt{unspecified}$ & $\texttt{indef}$ & $\texttt{limited}$ & $\texttt{stated}$ \\
\hlineB{2}
\begin{tabular}[l]{@{}l@{}}$\texttt{service}$ \\ $\texttt{security}$\end{tabular}
 & \begin{tabular}[l]{@{}l@{}}
\includegraphics[width=0.05\linewidth]{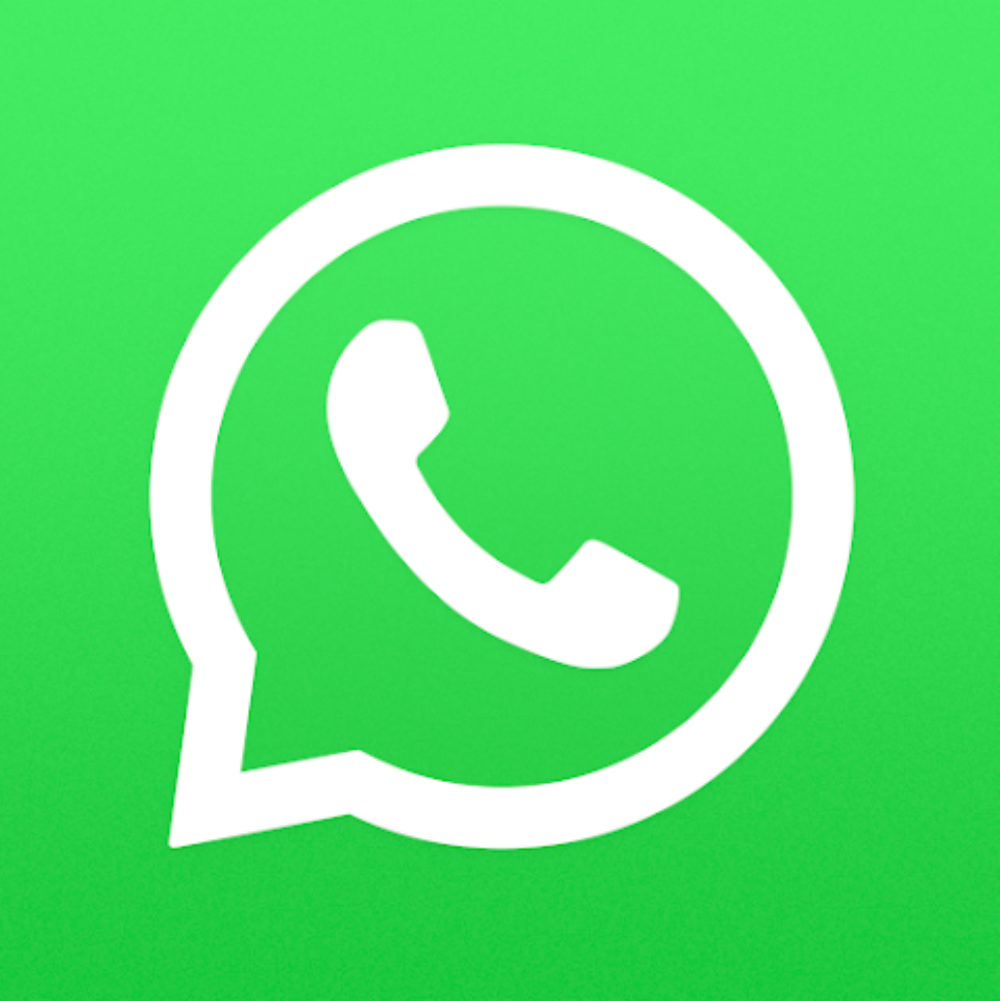}\includegraphics[width=0.05\linewidth]{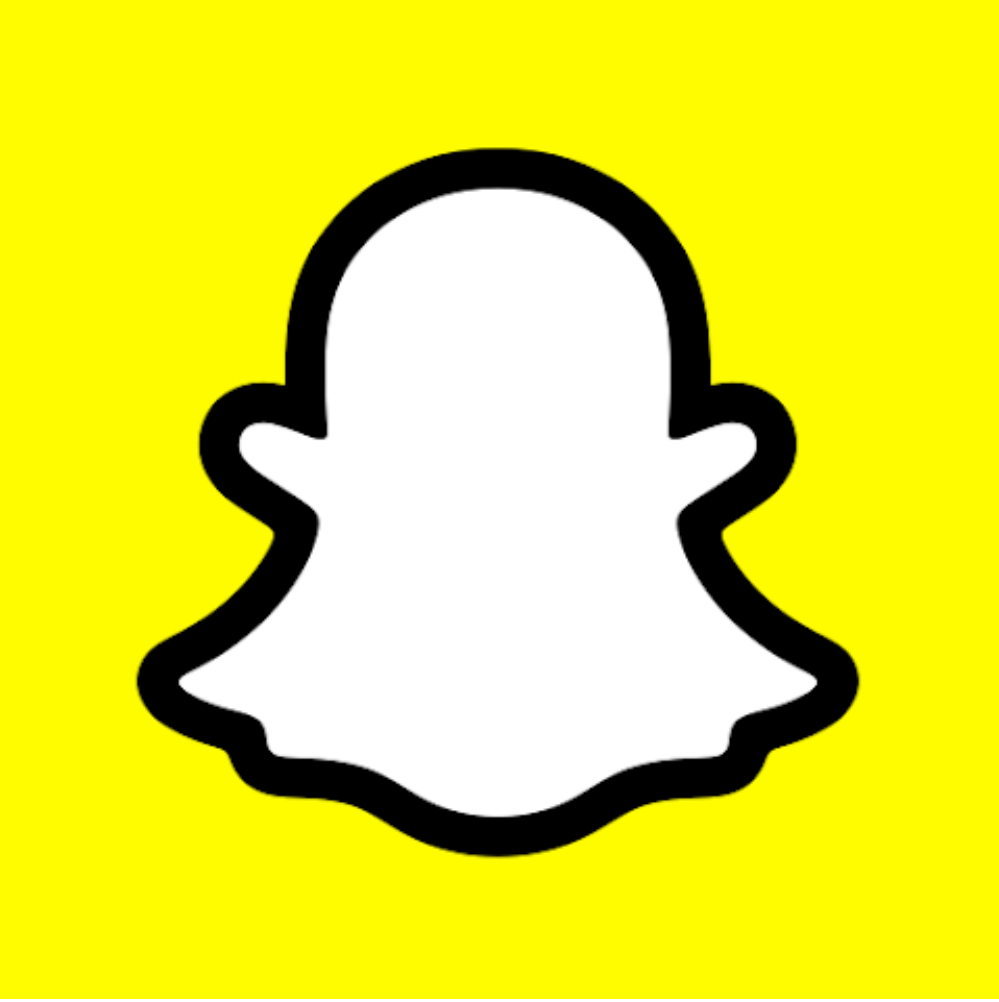}\includegraphics[width=0.05\linewidth]{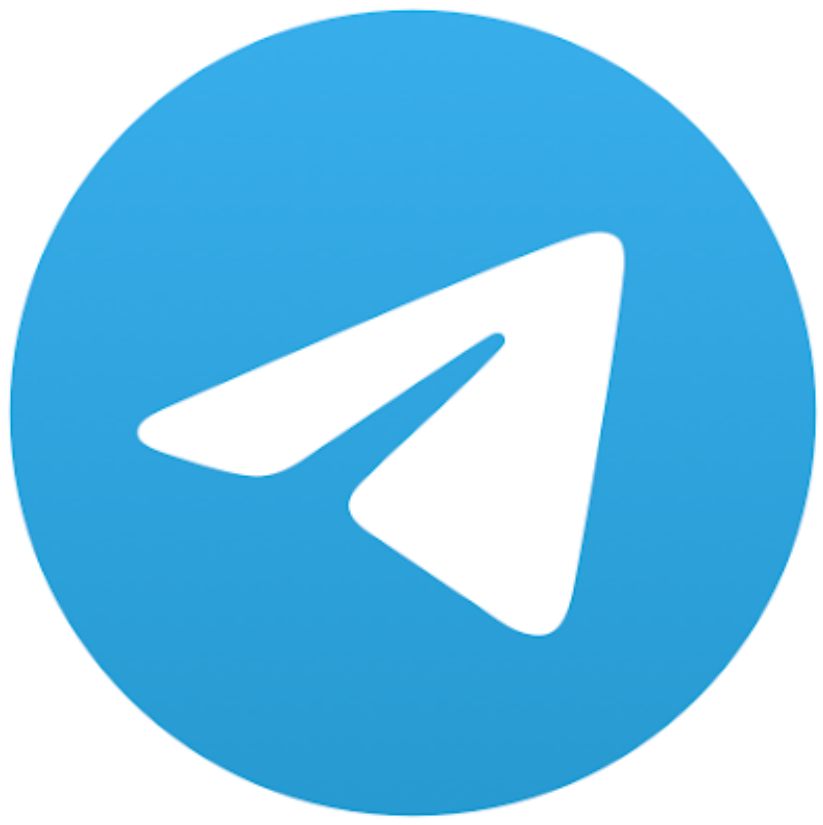}\\\includegraphics[width=0.05\linewidth]{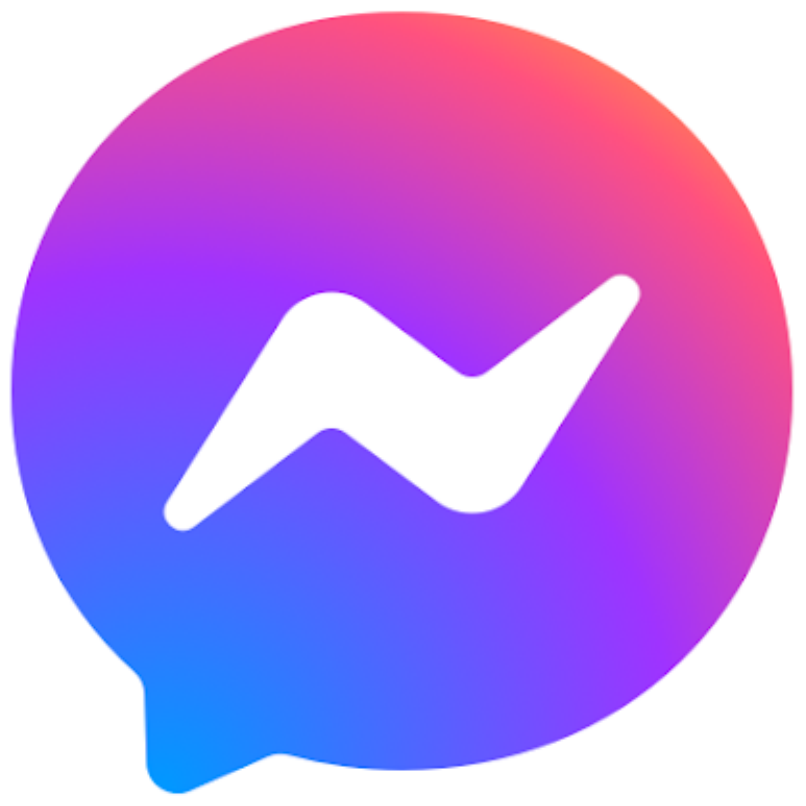}\includegraphics[width=0.05\linewidth]{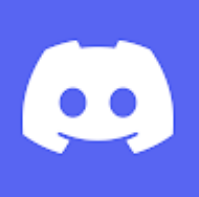}\includegraphics[width=0.05\linewidth]{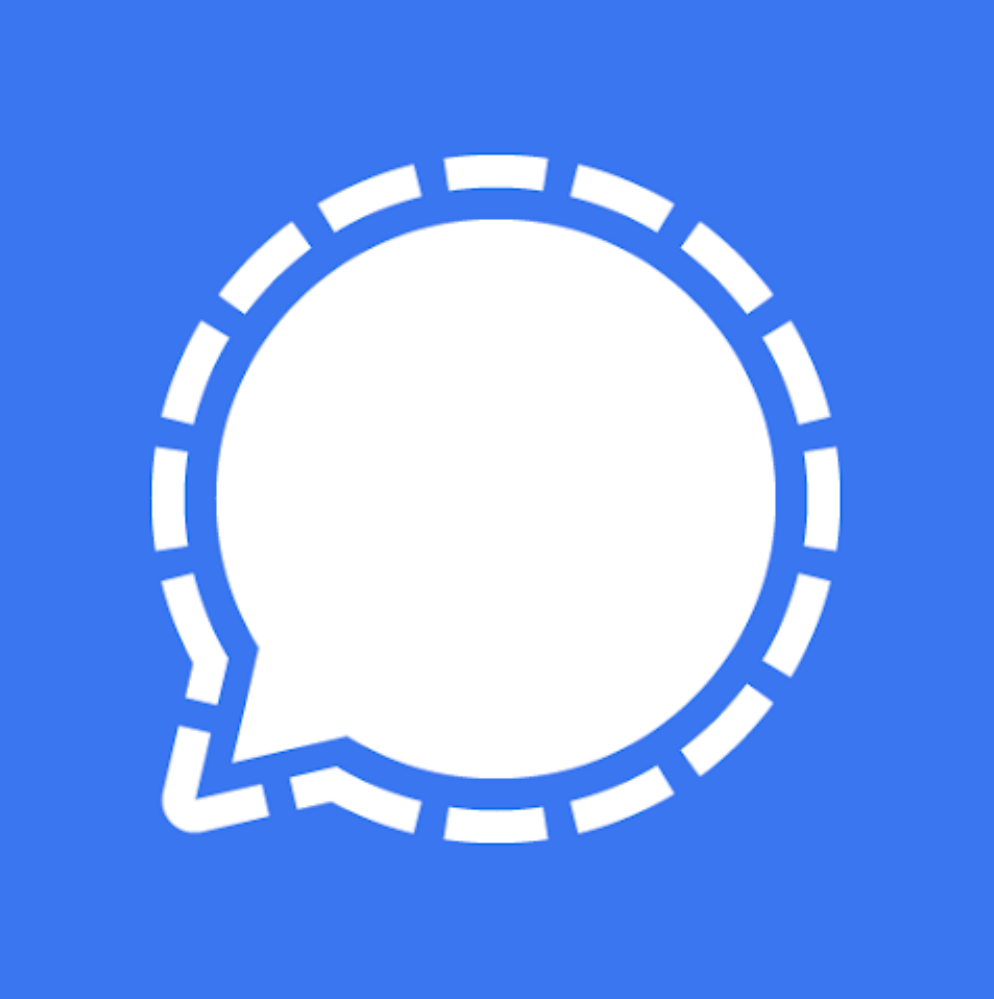} \end{tabular}&  &\includegraphics[width=0.05\linewidth]{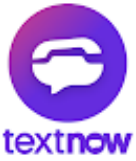} & \\
\hline
$\texttt{legal}$ &  & &\includegraphics[width=0.05\linewidth]{snapchat.pdf}\includegraphics[width=0.05\linewidth]{fb.pdf}\includegraphics[width=0.05\linewidth]{discord.pdf} &\includegraphics[width=0.05\linewidth]{whatsapp.pdf}\\
\hline
$\texttt{analytics}$ & \includegraphics[width=0.05\linewidth]{fb.pdf}\includegraphics[width=0.05\linewidth]{textnow.pdf} & &\includegraphics[width=0.05\linewidth]{whatsapp.pdf}\includegraphics[width=0.05\linewidth]{snapchat.pdf}\includegraphics[width=0.05\linewidth]{telegram.pdf}\includegraphics[width=0.05\linewidth]{discord.pdf} & \\
\hline
$\texttt{advertising}$ & \includegraphics[width=0.05\linewidth]{snapchat.pdf}\includegraphics[width=0.05\linewidth]{fb.pdf}\includegraphics[width=0.05\linewidth]{textnow.pdf} & &\includegraphics[width=0.05\linewidth]{whatsapp.pdf}\includegraphics[width=0.05\linewidth]{discord.pdf} & \\
\hline
$\texttt{unspecified}$ & \includegraphics[width=0.05\linewidth]{snapchat.pdf}\includegraphics[width=0.05\linewidth]{fb.pdf} & &
\begin{tabular}[l]{@{}l@{}}\includegraphics[width=0.05\linewidth]{whatsapp.pdf}\includegraphics[width=0.05\linewidth]{telegram.pdf}\includegraphics[width=0.05\linewidth]{textnow.pdf}\\\includegraphics[width=0.05\linewidth]{discord.pdf}\includegraphics[width=0.05\linewidth]{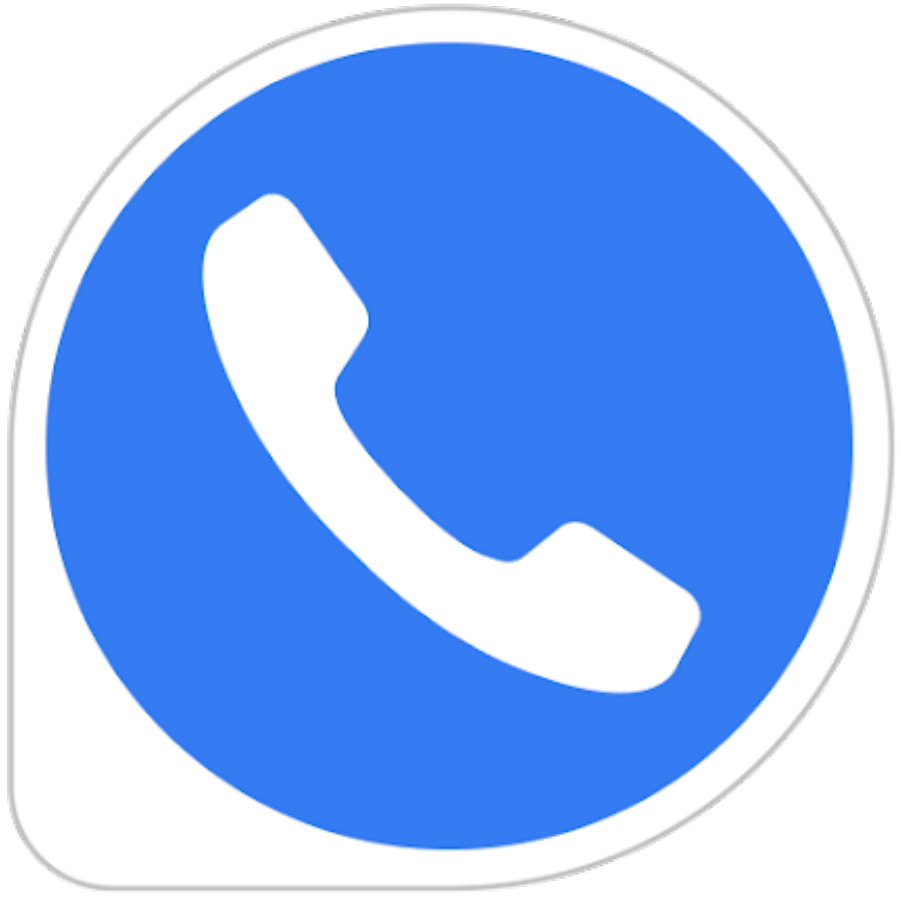}\includegraphics[width=0.05\linewidth]{signal.pdf}\end{tabular} & \\
\hlineB{2}
\end{tabular}
\end{center}
\caption{Privacy Comparison Shopping on Top Apps from the Communication Category. Each cell represents a specific purpose and retention-period setting. Such a fine-grained analysis can help users to choose applications based on their preferred privacy settings.}
\label{fig:privacy_shop}
\end{table}

We emphasize that existing sentence-classification-based approaches are not able to achieve this granularity as they classify these attributes individually. \name, on the other hand, provides a more fine-grained view by considering the joint distributions of these practices. For instance, individual analysis would describe \lstinline{retention-period} of \textit{Whatsapp} as \lstinline{limited}. However, this misses out on the information that data collected for specific purposes like \lstinline{perform-service} or \lstinline{security} is actually retained for an \lstinline{unspecified} period of time.

From \autoref{fig:privacy_shop}, we can observe the following trends, which can be useful for a user when comparing communication apps:
\begin{enumerate}
    \item \textit{Whatsapp}, \textit{Discord} and \textit{Signal} have better data retention practices as compared to \textit{Snapchat} and \textit{Messenger}.

    \item Only \textit{Textnow} specifies a \lstinline{retention-period} for the data retained for \lstinline{perform-service} or \lstinline{security} purposes. However, it is not transparent about data retained for \lstinline{advertising} or \lstinline{analytics}.

    \item None of the apps retain any data for an indefinite period of time.
\end{enumerate}

\section{Limitations}

\noindent \textbf{LLM and RAG Limitations.} 
\name uses both Large Language Models (LLMs) and the Retrieval-Augmented Generation (RAG) techniques to generate the truth table and inherits limitations associated with these. The entailment task employed relies on retrieving relevant context from a vector database of policy chunks (Figure.~\ref{fig:retrieval}). Errors within this retrieval pipeline can propagate throughout the system, potentially compromising the effectiveness of \name. Furthermore, entailment tasks populate the truth table, requiring LLMs to perform evidence-based reasoning. As we observed in \autoref{sec:tosdr_entail}, LLMs can exhibit nit-picky behavior, overlooking common contextual cues. Additionally, they can make erroneous inferences, as seen in  \autoref{sec:tosdr_entail}. Therefore, LLMs may struggle with the inherent ambiguity of natural language and misinterpret the meaning of contextual phrases or clauses within privacy policies. A potential mitigation strategy here could be to fine-tune LLMs, specifically on privacy entailment tasks, and use it as the compiler to generate the truth table. However, this approach requires significant training data and computational resources.

\noindent \textbf{Scaling number of atomic formulae.} The size and complexity of the atomic formulae in \name are dependent on the underlying privacy policy taxonomy. As the number of attributes within the taxonomy increases, the number of atomic formulae grows exponentially. This can lead to significant computational costs, potentially impacting the scalability of the framework. While this work focuses on providing a framework for downstream tasks given a pre-defined taxonomy, building an optimal taxonomy remains an open challenge. We acknowledge this limitation and emphasize that \name facilitates various applications once a suitable taxonomy is established. One potential solution is to upper bound the number of attributes that can be part of an atomic formula, thus reducing the cardinality of the joint distribution. For instance, consider a high-level category with $3$ attributes where each can assume $10$ different values. The cardinality of the joint distribution of all three attributes will be $1000$. However, if we only allow almost $2$ attributes in the joint distribution, we get $3$ different pairings among attributes, each with a cardinality of $100$. Overall, this reduces the number of rows in the truth table from $1000$ to $300$. This, however, gives a less fine-grained representation as compared to the $3$ attribute joint distribution. Therefore, this provides a way to balance efficiency and utility.

\noindent\textbf{Assumptions on Taxonomy.} For the truth table to serve as a comprehensive representation of a policy, the taxonomy must contain all relevant privacy-related topics. As seen in Section~\ref{sec:compliance}, incomplete taxonomies can lead to incomplete truth tables. In this case, certain compliance rules were not directly mapped to the taxonomy. Therefore, we were not able to check compliance for those rules. We reiterate that this work focuses on the framework's development and downstream applications, assuming a well-defined taxonomy. A potential mitigation strategy for incomplete taxonomies involves leveraging taxonomy completion tools~\cite{shi2024taxonomy}. These tools can automatically identify and fill in missing nodes within the taxonomy, potentially improving the comprehensiveness of the truth table and its efficacy in supporting downstream applications. 

\noindent \textbf{Vulnerability to policy poisoning attacks.}
Recent work has demonstrated that retrieval-augmented LLMs are vulnerable to poisoning attacks~\cite{chaudhari2024phantom,zhong2023poisoning,zou2024poisonedrag}. These attacks add carefully crafted triggers to the input documents that can lead to adversary-controlled misbehavior of both the embedding as well as the response LLM. These attacks can also be adapted to target systems that comprise multiple LLM instances~\cite{mangaokar2024prp}. Such an attack trigger, when added to a privacy policy document, can lead to incorrect formulae valuation and subsequently affect the downstream applications. Defense against such attacks that target ML models is still an open problem.

\section{Discussion}

\noindent \textbf{Modular Design.} \name has two main components: instantiating the grammar to get a set of atomic formulae and implementing a valuation function based on the NLI entailment task. In this paper, we initialize the grammar using the OPP-115 taxonomy and implement the valuation function using retrieval-augmented LLMs. However, both these components can also be performed using alternate implementations. \name's grammar can also be instantiated using other taxonomies like MAPP~\cite{arora2022tale} and Privacy label taxonomy~\cite{khandelwal2023overview}. Since \name automates the extraction of atomic formulae from any taxonomy, \name's grammar can be easily extended to incorporate new taxonomies as well as adapt to changes in existing ones. \name's valuation function is based on entailment, which is a well-studied NLI task. Therefore, it can alternatively be implemented using models that specialize in NLI tasks~\cite{wang2021entailment}. This also means that \name can benefit from any future advancements in the NLI.

\noindent \textbf{Privacy Policies and Source Code.} Prior work has studied the consistency between privacy policies and the source code implementation of their corresponding apps~\cite{slavin2016toward}. However, they require manually creating a mapping between code constructs and privacy concepts. Given the code understanding capability of LLMs, \name provides a way to automate this analysis.

\noindent \textbf{Beyond Privacy Policies.} \name's methodology can be used to compare any unstructured data, not just privacy policy documents. Given the relevant taxonomy, the \name framework can construct the corresponding atomic formulae and their valuations. Given the advancement in multi-modal LLMs, this can also include other modalities like vision and audio. Another interesting application of \name's consistency formulation is hallucination detection. Given a set of atomic formulae, an LLM's generated output should have valuations that are consistent with its input context. Any inconsistencies are likely the result of hallucinations or reasoning errors.

 \section{Conclusion}
In conclusion, \name advances automated privacy policy analysis by converting their complex text into a machine-readable format using valuations of atomic formulae. We implement a compiler for \name using off-the-shelf open-source LLMs and embedding models to evaluate a complex set of logical formulae based on the full text of a policy. This compiler achieves high precision and recall on the ToD;DR dataset. \name's applications in policy compliance, inconsistency detection, and privacy comparison shopping demonstrate its potential to make privacy policy analysis more accessible and understandable. 

\bibliography{example_paper}
\end{document}